\def\bey{\begin{eqnarray}}
\def\eey{\end{eqnarray}}
\def\be{\begin{equation}}
\def\ee{\end{equation}}
\def\ba{\begin{array}}
\def\ea{\end{array}}
\def\gm{\gamma}

\def\ld{\lambda}
\def\Ld{\Lambda}
\def\af{\alpha}
\def\sg{\sigma}

\def\om{\omega}
\def\r{\rho}

\def\bt{\beta}
\def\vep{\varepsilon}

\def\pp{\partial}
\def\pp{\partial}

\def\nnb{\nonumber}
\bigskip
\documentclass[twocolumn,showpacs,preprintnumbers,amsmath,amssymb,aps,prlamsfronts]{revtex4}
\usepackage{graphicx}
\usepackage{fancyhdr}
\usepackage{dcolumn}
\usepackage{bm}
\usepackage{amssymb}
\usepackage{amsmath}
\usepackage{graphicx}
\usepackage[normalem]{ulem}
\usepackage[dvips]{color}

\begin{document}
\preprint{ }
\title{ Effects of a weakly interacting light U boson  on the nuclear equation
of state and  properties of neutron stars in relativistic models}
\author{Dong-Rui Zhang$^1$, Ping-Liang Yin$^1$, Wei Wang$^1$, Qi-Chao Wang$^1$,
 Wei-Zhou Jiang$^{1,2}$\footnote{wzjiang@seu.edu.cn}}
\affiliation{  $^1$ Department of Physics, Southeast University,
Nanjing 211189, China\\ $^2$ National  Laboratory of Heavy Ion
Accelerator, Lanzhou 730000, China}

\begin{abstract}
 We investigate the effects of the light vector
U-boson that couples weakly to nucleons in relativistic mean-field
models on the equation of state and subsequently the consequence in
neutron stars. It is analyzed that the U-boson can lead to a much
clearer rise of the neutron star maximum mass in models with the much
softer equation of state. The inclusion of the U-boson may thus allow
the existence of the non-nucleonic degrees of freedom in the interior
of large mass neutron stars initiated  with the favorably soft EOS of
normal nuclear matter.  In addition, the sensitive role of the
U-boson in the neutron star radius  and its relation to the test of
the non-Newtonian gravity that is herein addressed by the light
U-boson are discussed.

\end{abstract}
 \maketitle

\section{Introduction}
Confronting nuclear physics, we should highlight the great importance
of the equation of state (EOS), for it being significantly important
to study the structure of nuclei, the reaction dynamics of heavy-ion
collisions, and many issues in
astrophysics~\cite{lat00,Hor01,Ste05,Li08}. The nuclear EOS consists
usually of two ingredients: the energy density for symmetric matter
and the density dependence of the symmetry energy. For the former,
the saturation properties are quite clear nowadays, though its
high-density behavior remains to be revealed in more details.
However, the density dependence of the symmetry energy is still
poorly known especially at high
densities~\cite{Li08,Brown00,Ku03,Die03}, and even the trend of the
density dependence of the symmetry energy  can be predicted to be
contrary. While most relativistic
theories~\cite{Hor01,Ste05,Lee98,to03,ji05,ji07,Chen07} and some
non-relativistic theories~\cite{Brown00,Die03,Chen05,Li06} predict
that the symmetry energy increases continuously at all densities,
many other  non-relativistic theories (for instance, see
~\cite{Brown00,St03,Chen05,Roy09}), in contrast, predict that the
symmetry energy first increases, then decreases above certain
supra-saturation densities, and even in some
predictions~\cite{Li08,Brown00,Ku03} becomes negative at high
densities, referred as the super-soft symmetry energy. Therefore, the
experimental extraction is of necessity.

Recently, by analyzing the FOPI/GSI  data on the $\pi^{-}/\pi^{+}$
radio in relativistic heavy-ion collisions~\cite{Re07}, the evidence
for a super-soft symmetry energy was found~\cite{Xiao09}. This
finding can result in many consequences, while a direct challenge is
how to stabilize a normal neutron star with the super-soft symmetry
energy. Conventionally, a mechanical instability may occur if the
symmetry energy starts decreasing quickly above the certain
supra-saturation density~\cite{St03,Gle00,Wen09}.  To solve this
problem, one possible way is to take into account the hadron-quark
phase transition which lifts up the pressure in pure quark
matter~\cite{Al05}, while the transition is expected to occur at much
higher densities within a narrow region of parameters. Instead, one
may consider the possible correction to the gravity. Though the
gravitational force was first discovered in the history, it is still
the most poorly characterized, compared to three other fundamental
forces that can be favorably unified within the gauge theory. For the
further grand unification of four forces, the correction to the
conventional gravity seems necessary. The light U-boson, which is
proposed beyond the standard model, can play the role in deviating
from the inverse square law of the gravity due to the Yukawa-type
coupling, see Refs.~\cite{Wen09,Ad03,Kr09,Re09} and references
therein. This light U-boson was used as the interaction propagator of
the MeV dark matter and was used to account for the bright 511 keV
$\gamma$-ray from the galactic
bulge~\cite{Bo04,Boe04,Bor06,Zhu07,Fa07,Je03}. As a consequence of
its weak coupling to baryons, the stable neutron star can be obtained
in the presence of the super-soft symmetry energy~\cite{Wen09}. In
addition, it is noted that through the reanalysis of the FOPI/GSI
data with a different dynamical model another group extracted a
contrary density dependent trend of the symmetry energy at high
densities~\cite{Fen10}. The solution of the controversy is still in
progress.

In pursuit of the covariance in addressing neutron stars bound by the
strong gravity, the relativistic models are favorable to obtain the
EOS, though the fraction, arisen from the relativistic effect of fast
particles in the compact core of neutron stars, is just moderate.
Apart from the non-relativistic models to obtain the EOS of neutron
stars in Ref.~\cite{Wen09}, we will adopt the relativistic mean-field
(RMF) models in this work. The RMF theory which is based on the Dirac
equations for nucleons with the potentials given by the meson
exchanges achieved great success in the past few
decades~\cite{Wal74,Bog77,Chin77,Ser86,Rei89,Ring96,Ser97,Ben03,Meng06,Ji07}.
The original Lagrangian of the RMF model was first proposed by
Walecka more than 30 years ago~\cite{Wal74}. The Walecka model and
its improved versions were characteristic of the cancellation between
the big attractive scalar field and the big repulsive vector field.
To soften the EOS obtained with the simple Walecka model, the proper
medium effects were accounted with the inclusion of the nonlinear
self-interactions of the $\sigma$ meson proposed by Boguta et.
al.~\cite{Bog77}. A few successful nonlinear RMF models, such as
NL1~\cite{Re86}, NL2~\cite{Lee86}, NL-SH~\cite{Sha93},
NL3~\cite{La97}, and etc., had been obtained by fitting saturation
properties and ground-state properties of a few spherical nuclei.
Later on, an extension to include the self-interaction of $\omega$
meson was implemented to obtain RMF potentials which were required to
be consistent with the Dirac-Brueckner self-energies~\cite{Su94}. In
this direction, besides the early model TM1~\cite{Su94}, there were
recent versions PK1~\cite{Lo04} and FSUGold~\cite{Pie05}.

Although various RMF models reproduce successfully the saturation
properties of nuclear matter and structural properties of finite
nuclei, the corresponding EOS's  may behave quite differently at high
densities especially in isospin-asymmetric nuclear matter. It was
reported in the literature~\cite{Kr09,Wen09} that the light U-boson
can significantly modify the EOS in isospin-asymmetric matter.
However, the further systematic work to analyze the effect of the
light U-boson on various nuclear EOS's is still absent. In this work,
we will investigate in detail the effect of light U-boson on the EOS
and properties of neutron stars with various RMF models. In
particular, we will address the difference of the effects induced by
the U-boson in various RMF models.

The paper is organized as follows. In Sec.~\ref{rmf}, we present
briefly the formalism based on the Lagrangian of the relativistic
mean-field models. In Sec.~\ref{results}, numerical results and
discussions are presented. At last, a summary is given in
Sec.~\ref{summary}.

\section{Formalism}
\label{rmf} In the RMF approach, the nucleon-nucleon interaction is
usually described via the exchange of three mesons: the isoscalar
meson $\sg$, which provides the medium-range attraction between the
nucleons, the isoscalar-vector meson $\om$, which offers the
short-range repulsion, and the isovector-vector meson $b_0$, which
accounts for the isospin dependence of the nuclear force. The
relativistic Lagrangian can be written as: \bey
 {\cal L}&=&
{\overline\psi}[i\gm_{\mu}\partial^{\mu}-M+g_{\sg}\sg-g_{\om}
\gm_{\mu}\om^{\mu}-g_\r\gm_\mu \tau_3 b_0^\mu
   ]\psi\nnb\\
      &  &
    - \frac{1}{4}F_{\mu\nu}F^{\mu\nu}+
      \frac{1}{2}m_{\om}^{2}\om_{\mu}\om^{\mu}
    - \frac{1}{4}B_{\mu\nu} B^{\mu\nu}
     +\frac{1}{2}m_{\r}^{2} b_{0\mu} b_0^{\mu}\nnb
    \\
     &&+
\frac{1}{2}(\partial_{\mu}\sg\partial^{\mu}\sg-m_{\sg}^{2}\sg^{2})
+U_{\rm eff}(\sg,\om, b_0)+{\cal L}_u, \label{eq:lag1}
  \eey
 where
 $\psi,\sg,\om$,$b_0$ are the fields of
the nucleon, scalar, vector, and neutral isovector-vector mesons,
with their masses $M, m_\sg,m_\om$, and $m_\r$, respectively.
$g_i(i=\sg,\om,\r)$  are the corresponding meson-nucleon
couplings. $F_{\mu\nu}$ and $ B_{\mu\nu}$ are the strength tensors
 of $\om$ and $\r$ mesons respectively,
\begin{equation}\label{strength} F_{\mu\nu}=\pp_\mu
\om_\nu -\pp_\nu \om_\mu,\hbox{  } B_{\mu\nu}=\pp_\mu b_{0\nu}
-\pp_\nu b_{0\mu}.
\end{equation}
The self-interacting terms of $\sigma$, $\om$ mesons and  the
isoscalar-isovector coupling  are given generally as
 \bey
 U_{\rm eff}(\sg,\om^\mu, b_0^\mu)&=&-\frac{1}{3}g_2\sg^3-\frac{1}{4}g_3\sg^4
 +\frac{1}{4}c_3(\om_\mu\om^\mu)^2\nnb\\
 &&+4\Ld_{V}g^2_\r  g_\om^2
 \om_\mu\om^\mu b_{0\nu}b_0^\nu. \label{eq:u}
 \eey
Here, the isoscalar-isovector coupling term is introduced to
modify the density dependence of the symmetry energy~\cite{Hor01}.
In addition, we include in Lagrangian ${\cal L}_u$ for the U-boson
that is beyond the standard model. A very light U-boson can be
utilized to interpret the deviation from the Newton's
gravitational potential which is usually characterized in the
form~\cite{Wen09,Kr09}:
   \be V(r)=-\frac{G_\infty m_1m_2}{r}(1+\af e^{-r/\ld}) \label{grav}\ee
where $G_\infty$ is the universal gravitational constant,
$\af=-g^2_u/4\pi G_\infty M_B^2$ is a dimensionless strength
parameter with $g_u$ and $M_B$ being the boson-nucleon coupling
constant and baryon mass, respectively, and $\ld=1/m_u$ is the length
scale with $m_u$ being the boson mass. According to the conventional
view, the Yukawa-type correction to the Newtonian gravity resides at
the matter part rather than the geometric part. Thus, following the
form of the vector meson, ${\cal L}_u$  is written as:
 \bey {\cal L}_u&=&-{\overline\psi}g_u\gm_{\mu}u^{\mu}\psi-\frac{1}{4}U_{\mu\nu} U^{\mu\nu}
      +\frac{1}{2}m_u^{2}u_{\mu}u^{\mu},
      \eey
with $u$ the field of U-boson. $U_{\mu\nu}$ is the strength tensor
of U-boson,
\begin{equation} U_{\mu\nu}=\pp_\mu u_{\nu} -\pp_\nu u_{\mu}.
\end{equation}

With the standard Euler-Lagrange formala, we can deduce from the
Lagrangian the equations of motion for the nucleon and mesons. They
are given as follows: \be
[i\gm_{\mu}\partial^{\mu}-M+g_{\sg}\sg-g_{\om}
\gm_{\mu}\om^{\mu}-g_u\gm_{\mu}u^{\mu}-g_\r\gm_\mu \tau_3 b_0^\mu
   ]\psi=0
   \ee
   \bey
   (\partial_t^2-\bigtriangledown^2+m_\sg^2)\sg&=&g_\sg{\overline\psi}\psi-g_2\sg^2-g_3\sg^3,\\
   (\partial_t^2-\bigtriangledown^2+m_\om^2)\om_{\mu}&=&g_\om{\overline\psi}\gm_{\mu}\psi-
              c_3\om_{\mu}^3\nnb\\
              && -8\Ld_{V} g_\r^2g_\om^2b_{0\nu} b_0^{\nu}\om_\mu,\\
   (\partial_t^2-\bigtriangledown^2+m_\r^2)b_{0\mu}&=&g_\r{\overline\psi} \gm_{\mu}
   \tau_3 \psi\nnb\\
   &&-8\Ld_{V} g_\r^2g_\om^2 \om_\nu\om^\nu b_{0\mu},\\
   (\partial_t^2-\bigtriangledown^2+m_u^2)u_\mu&=&g_u{\overline\psi}\gm_{\mu}\psi.
 \eey In the mean-field approximation, all
derivative terms drop out and the expectation values of space-like
components of vector fields vanish (only zero components survive) due
to translational invariance and rotational symmetry of the nuclear
matter. In addition, only the third component of isovector fields
survives because of the charge conservation. In the mean-field
approximation, after the Dirac field of nucleons is
quantized~\cite{Ser86},  the fields of mesons and U-boson, which are
replaced by their classical expectation values, obey following
equations: \bey
m_\sg^2\sg&=&g_\sg\r_s-g_2\sg^2-g_3\sg^3,\\
   m_\om^2\om_0&=&g_\om\r_B-c_3\om_0^3-8\Ld_{V} g_\r^2g_\om^2b_0^2\om_0,\\
   m_\r^2b_0&=&g_\r \r_3-8\Ld_{V} g_\r^2g_\om^2 \om_0^2b_0,\\
   m_u^2u_0&=&g_u\r_B,
\eey where $\r_s$ and $\r_B$ are the scalar and baryon densities,
respectively, and $\r_3$ is the difference between the proton and
neutron densities, namely, $\r_3=\r_p-\r_n$.  The set of coupled
equations can be solved self-consistently using the iteration method.
With these mean-field quantities, the resulting   energy density
$\vep$ and pressure $P$ are written as:
 \bey
 \vep&=&\sum_{i=p,n}\frac{2}{(2\pi)^3}\int^{k_{F_i}} d^3\!k E^*_i+
 \frac{1}{2}m_\om^2\om_0^2+\frac{1}{2}\frac{g_u^2}{m_u^2}\r_B^2\nnb\\
 && + \frac{1}{2}m_\sg^2\sg_0^2+\frac{1}{2}m_\r^2 b_0^2
 +\frac{1}{3}g_2\sg^3+\frac{1}{4}g_3\sg^4\nnb\\
&&+\frac{3}{4}c_3\om_0^4+12\Ld_{V}g^2_\r g_\om^2
 \om_0^2 b_0^2,\label{eq:e}\\
P&=&\frac{1}{3}\sum_{i=p,n}\frac{2}{(2\pi)^3}\int^{k_{F_i}} d^3\!k
\frac{{\bf k}^2}{E^*_i}+\frac{1}{2}m_\om^2\om_0^2+\frac{1}{2}
\frac{g_u^2}{m_u^2}\r_B^2\nnb\\
&&-\frac{1}{2}m_\sg^2\sg_0^2+\frac{1}{2}m_\r^2 b_0^2
-\frac{1}{3}g_2\sg^3-\frac{1}{4}g_3\sg^4\nnb\\
&& +\frac{1}{4}c_3\om_0^4+4\Ld_{V}g^2_\r g_\om^2
 \om_0^2 b_0^2, \label{eq:p}\eey
with $E^*_i=\sqrt{{\bf k}^2+(M^*_i)^2}$.

Given above is the formalism for nuclear matter without considering
the $\beta$ equilibrium. For asymmetric nuclear matter at $\beta$
equilibrium, the chemical equilibrium and charge neutrality
conditions need to be additionally considered, which are written as:
 \bey
  \mu_n&=&\mu_p+\mu_e,\\
  \r_e&=&\r_p,\\
  \r_B&=&\r_n+\r_p,
 \eey
where $\mu_n, \mu_p, \mu_e$ are the chemical potential of neutron,
proton and electron, respectively, and $\r_e$ is the number density
of electrons. In neutron star matter, the EOS is obtained by adding
in Eqs.(\ref{eq:e}) and (\ref{eq:p}) the contribution of the free
electron gas.

The neutron star properties are obtained from solving the
Tolman-Oppenheimer-Volkoff (TOV) equation~\cite{Op39,Tol39}:
\begin{eqnarray}
\frac{dP(r)}{dr}&=&-\frac{[P(r)+\vep(r)][M(r)+4\pi r^3 P(r)]}
{r(r-2M(r))},\label{eq:tov1}\\
M(r)&=&4\pi\int^r_0d\!\tilde{r}\tilde{r}^2
\vep(\tilde{r}),\label{eq:tov2}
\end{eqnarray}
where $r$ is the radial coordinate from the center of the star,
$P(r)$ and $\vep(r)$ are the pressure and energy density at position
$r$, respectively, and $M(r)$ is the mass contained in the sphere of
the radius $r$. Note that here we use units for which the gravitation
constant is $G_\infty=c=1$. The radius $R$ and mass $M(R)$ of a
neutron star are obtained from the condition $p(R)=0$. Because the
neutron star matter, consisting of neutrons, protons, and electrons
(npe) at $\bt$ equilibrium in this work, undergoes a phase transition
from the homogeneous matter to the inhomogeneous matter at the low
density region, the RMF EOS obtained from the homogeneous matter does
not apply to the low density region. For a thorough description of
neutron stars, we thus adopt the empirical low-density EOS in the
literature~\cite{Ba71,Ii97}.

\section{Results and discussions}
\label{results}

Among a number of nonlinear RMF parametrizations, we select several
typical best-fit parameter sets, for instance NL1~\cite{Re86},
NL-SH~\cite{Sha93}, NL3~\cite{La97}, TM1~\cite{Su94} and
FSUGold~\cite{Pie05}, to investigate the effects of the U-boson on
the EOS of isospin-asymmetric nuclear matter and properties of
neutron stars. The nonlinear RMF models usually include the nonlinear
self-interactions of the $\sigma$ meson to simulate appropriate
medium dependence of the strong interaction. This is typical in RMF
parameter sets NL1, NL-SH and NL3. In addition to the nonlinear
$\sigma$ meson self-interactions, in TM1 and FSUGold the nonlinear
self-interaction of the $\omega$ meson is also included. Parameters
and saturation properties of these parameter sets are listed in
Table~\ref{t:t1}.
  \begin{table*}
  \begin{center}
 \caption{Parameters and saturation properties for various parameter
sets. Here, the NL3$\Ld_V$ is the same as the original parameter set
NL3 but with the readjusted $g_\rho$ after the $\Ld_V$ is included to
modify the density dependence of the symmetry energy, and the
TM1$\Ld_V$ to the TM1 is the same as the NL3$\Ld_V$ to the NL3. Meson
masses, incompressibility and symmetry energy are in units of MeV,
and the density is in unit of $fm^{-3}$. \label{t:t1}}
    \begin{tabular}{ c c c c c c c c c c c c c c c}
\hline\hline &$g_\sg$ & $g_\om$& $g_\r$& $m_\sg$ & $m_\om$  &
$m_\rho$ & $g_2$ & $g_3$& $c_3$ & $\Ld_V$ & $\rho_0$ &
$\kappa$ & $M^*/M$ & $E_{sym}$ \\
\hline
NL1     &10.138 &13.285 &4.976 &492.250 &795.359 &763 &12.172 &-36.265 &- &-&0.153 &211.3 &0.57 &43.7 \\
NL-SH   &10.444 &12.945 &4.383 &526.059 &783.000 &763 &6.910 &-15.834 &- &- &0.146 &355.4 &0.60 &36.1\\
NL3     &10.217 &12.868 &4.474 &508.194 &782.501 &763 &10.431 &-28.890  &- &- &0.148 &271.8 &0.60 &37.4\\
TM1     &10.029 &12.614 &4.632 &511.198 &783.000 &770 &7.233 &0.618 &71.31&- &0.145 &281.2 &0.63 &36.9\\
FSUGold &10.592 &14.302 &5.884 &491.500 &782.500 &763 &4.277 &49.934 &418.39 &0.03 &0.148 &230.0 &0.61 &32.5\\
NL3$\Ld_V$&10.217 &12.868 &5.664 &508.194 &782.501 &763 &10.431 &-28.890  &- &0.03 &0.148 &271.8 &0.60 &31.8\\
TM1$\Ld_V$&10.029 &12.614 &5.720 &511.198 &783.000 &770 &7.233 &0.618 &71.31&0.03 &0.145 &281.2 &0.63 &32.1\\

\hline\hline
\end{tabular}
\end{center}
\end{table*}

\begin{figure}[thb]
\begin{center}
\vspace*{-5mm}
\includegraphics[height=8.5cm,width=8.5cm]{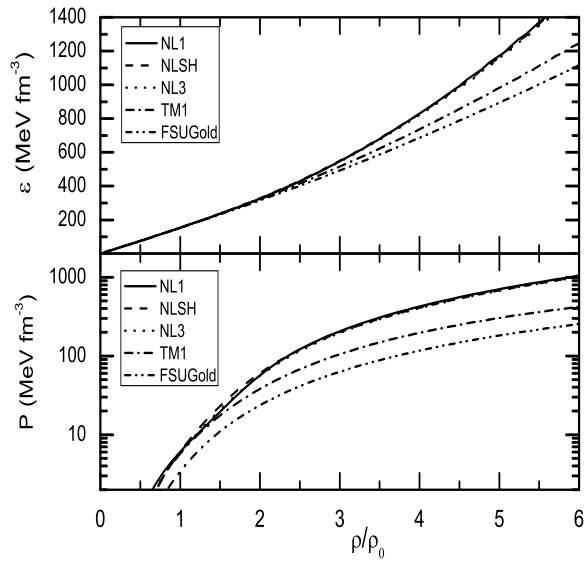}
 \end{center}
\vspace*{-5mm}\caption{Energy density $\vep$ (upper panel) and
pressure P (lower panel) as a function of density with various RMF
parameter sets, NL3, NL1, NL-SH, TM1, and FSUGold in npe matter at
$\beta$ equilibrium. \label{f:rho}}
\end{figure}

In Fig.~\ref{f:rho}, the energy density and pressure of npe matter at
$\beta$ equilibrium are shown as a function of nucleon density for
various models without the inclusion of the U-boson. It is seen that
the EOS with parameter sets TM1 and FSUGold is clearly softer than
that with the NL1, NL-SH and NL3 with the increase of the density.
The softening stems from the inclusion of the nonlinear
self-interaction of the $\om$ meson that lowers the repulsion
provided by the $\omega$ meson at high densities, while the excess
softening with the FSUGold as compared to that with the TM1 can be
attributed dominately to the larger parameter $c_3$ in FSUGold.

\begin{figure}[thb]
\begin{center}
\vspace*{-5mm}
\includegraphics[height=8.5cm,width=8.5cm]{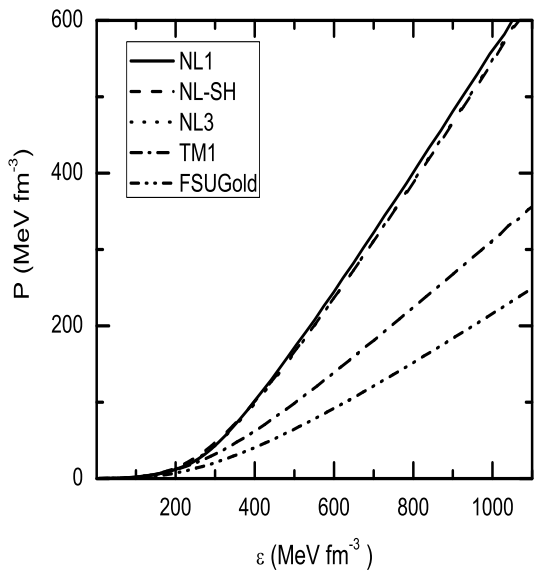}
 \end{center}
\vspace*{-5mm}\caption{The correlation between the pressure and the
energy density  in npe matter at $\beta$ equilibrium with various RMF
models. \label{ep0}}
\end{figure}
Shown in Fig.~\ref{ep0} is the correlation between the pressure and
the energy density given in Fig.~\ref{f:rho}. This correlation is
usually regarded as the EOS that is used as the input of the
Tolman-Oppenheimer-Volkoff (TOV) equation~\cite{Op39,Tol39} for the
evaluation of the neutron star properties. Once again, we see the
large deviations in the EOS with different RMF models especially at
high densities.   In the following, it is thus interesting to see how
the U-boson affects the EOS produced by various RMF models that
differs largely at high densities.

\begin{figure}[thb]
\begin{center}
\vspace*{-5mm}
\includegraphics[height=10.2cm,width=8.5cm]{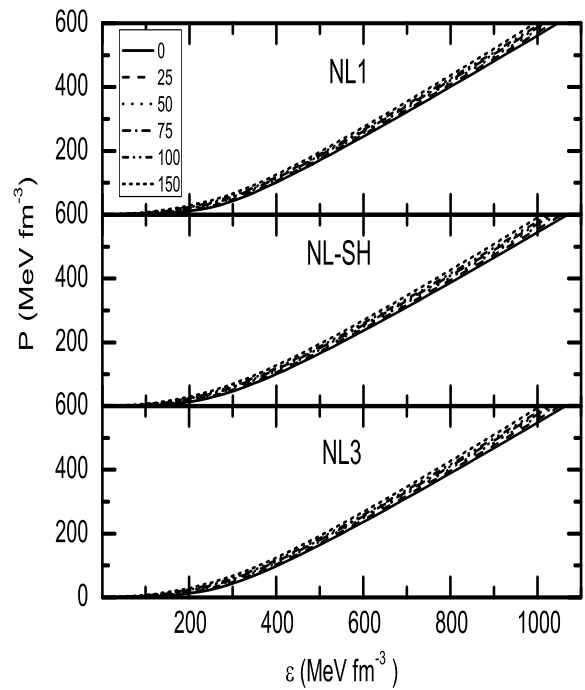}
 \end{center}
\vspace*{-5mm}\caption{Equation of state of neutron star matter with
three RMF models, NL3, NL1 and NL-SH with the inclusion of  the
U-boson. The numbers in the legend are the values of $(g_u/m_u)^2$ in
units of $GeV^{-2}$. \label{epnl}}
\end{figure}

\begin{figure}[thb]
\begin{center}
\vspace*{-5mm}
\includegraphics[height=8.5cm,width=8.5cm]{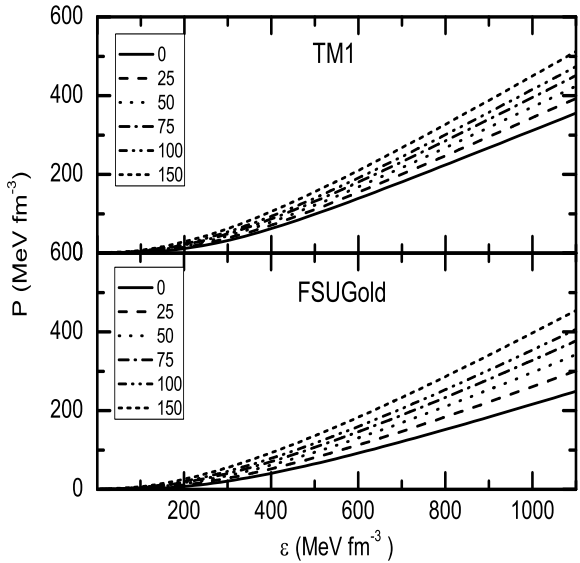}
 \end{center}
\vspace*{-5mm}\caption{The same as in Fig.~\ref{epnl} but for the RMF
models TM1 and FSUGold.\label{eptf}}
\end{figure}

In the RMF approximation, the contribution of the U-boson in a linear
form is just decided by the ratio of the coupling constant to its
mass, i.e., $g_u/m_u$, as seen in Eqs.(\ref{eq:e}) and (\ref{eq:p}).
In Figs.~\ref{epnl} and \ref{eptf}, the EOS's with various models are
depicted for a set of ratios $(g_u/m_u)^2$. It is shown in
Figs.~\ref{epnl} and \ref{eptf} that the inclusion of the U-boson
stiffens the EOS. This is physically obvious since the vector form of
the U-boson provides an excess repulsion in addition to the vector
mesons, whereas an interestingly large difference appears for
different types of models.  As shown in Figs.~\ref{epnl} and
\ref{eptf}, the EOS's with the TM1 and FSUGold acquires a much more
apparent stiffening than that with the NL1, NL-SH and NL3 by
including the U-boson. This phenomenon can be understood by the
inherent feature of these models. In models NL1, NL-SH and NL3, the
repulsion is quadratic in the density because the nonlinear
self-interaction of the $\om$ meson is not considered. With the
increase of the density, the repulsion provided by the $\omega$ meson
dominates the attraction provided by the $\sigma$  meson. The
cancellation between the repulsion and attraction in the pressure
(see Eq.(\ref{eq:p}) is not prominent at high densities so that the
U-boson plays a similar role in the energy density and pressure.
Thus, these EOS's are just moderately modified by the U-boson, as
shown in Fig.~\ref{epnl}. For models TM1 and FSUGold that feature a
clearly softer EOS at high densities, the cancellation between the
repulsion and attraction becomes significant and thus sharpens the
importance of the U-boson in the pressure. Comparing to the addition
of the big repulsion and attraction in the energy density, the
U-boson just plays a marginal role in modifying the energy density.
Thus, the U-boson can modify appreciably the correlation between the
pressure and energy density in the high-density region in favorably
softened models, for instance, the TM1 and FSUGold, as shown in
Fig.~\ref{eptf}. Because in TM1 and FSUGold the nonlinear term of the
$\om$ meson plays a decisive role in softening the EOS, the larger
the parameter $c_3$, the more apparent the modification, as shown
comparatively in the upper and lower panels of Fig.~\ref{eptf}.

\begin{figure}[thb]
\begin{center}
\vspace*{-5mm}
\includegraphics[height=8.5cm,width=8.5cm]{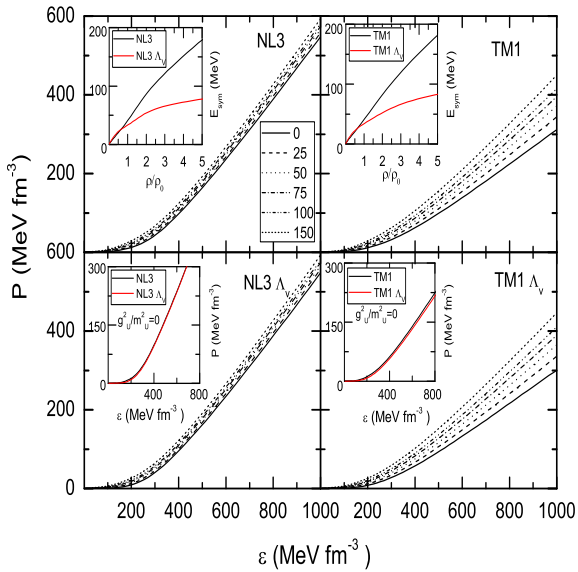}
 \end{center}
\vspace*{-5mm}\caption{(Color online) The same as in Fig.~\ref{epnl}
but to exhibit the difference between the cases with and without the
modification to the symmetry energy. Left panels represent the
results with the NL3 and NL3$\Ld_V$, and right panels are the results
with the TM1 and TM1$\Ld_V$. Different density dependencies of the
symmetry energy are drawn in the insets of upper panels, while given
in the insets of lower panels are the EOS of two cases in the absence
of the U-boson. \label{epnl3}}
\end{figure}

In addition, it is interesting to examine whether the significant
difference in the U-boson-induced modification to the EOS can be
created by softening the symmetry energy. The symmetry energy is
softened by including the isoscalar-isovector coupling term in RMF
models (see Eq.(\ref{eq:u})). In Fig.~\ref{epnl3}, we depict the EOS
without (upper panels) and with (lower panels) the softening of the
symmetry energy in NL3 and TM1. However, no visible difference in two
cases with the NL3 is observed, and with the TM1 the difference is
not significant. This observation seems to show a contrast with that
in Ref.~\cite{Wen09} where the fluffy EOS due to the super-soft
symmetry energy can be lifted up by the U-boson to support a normal
neutron star.  In deed, the magnitude of the modification to the EOS
caused by the U-boson relies on the softness of the EOS. As long as
the EOS is modified significantly by softening the symmetry energy,
the stiffening role of the U-boson in the EOS can be considerably
enhanced accordingly. Given that the stiff EOS with the NL3 is little
modified by softening the symmetry energy, as shown in the inset of
the left lower panel in Fig.~\ref{epnl3}, the softening of the
symmetry energy can scarcely affect the role of the U-boson. For
models with a softer EOS, the situation can turn out to be different
when the EOS is modified appreciably by softening the symmetry
energy. Indeed, the vital role of the U-boson in the EOS of the
non-relativistic MDI model with a super-soft symmetry
energy~\cite{Wen09} is a typical case that the role of the U-boson
can be largely amplified due to the softening of the symmetry energy.
In RMF models, for instance, the TM1 whose EOS is softer than that
with the NL3, the softening of the symmetry energy can also result in
some visible difference in the EOS and thereby the role of the
U-boson, as shown in right panels of Fig.~\ref{epnl3}.

\begin{figure}[thb]
\begin{center}
\vspace*{-5mm}
\includegraphics[height=8cm,width=8.5cm]{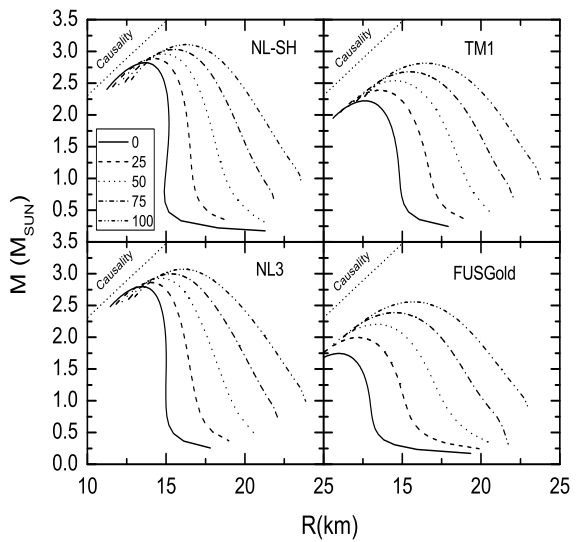}
 \end{center}
\vspace*{-5mm}\caption{The mass-radius relation of neutron stars with
various models. The U-boson is included with  various ratio
parameters of $(g_u / m_u)^2$.\label{mr}}
\end{figure}

Next, we turn to the consequences in hydrostatic neutron stars with
the EOS modified by the U-boson. Using Eqs.(\ref{eq:tov1}) and
(\ref{eq:tov2}), the mass and radius of  hydrostatic neutron stars
can be obtained with the given EOS. In Fig.~\ref{mr}, the mass-radius
(M-R) relation of neutron stars is depicted with different ratio
parameter $(g_u/m_u)^2$ for the U-boson in various models. With the
inclusion of the U-boson, we can see that both the maximum mass and
radius of neutron stars increase significantly. It is clearly seen
that the star maximum mass with the soft EOS is modified more
significantly by the U-boson. This is consistent with the
corresponding modification to the high-density EOS caused by the
U-boson, as shown in Figs.~\ref{epnl} and \ref{eptf}. The consistency
is established on the fact that the maximum mass of neutron stars is
dominated by the high-density behavior of the EOS. In the past, a few
neutron stars with large masses around $2M_\odot$ had been
observed~\cite{Ni05,Ni08,Oz06}. Though it can have improvements in
experimental aspects, the observation of neutron stars with large
masses is not so scarce. Recently, the mass of the LMXB 4U1608-52 is
measured to be 1.74$M_\odot$~\cite{Gu10}, and most recently a
$2M_\odot$ neutron star J1614-2230 was measured through the Shapiro
delay~\cite{De10}. Note that the model FSUGold which is well
consistent with the nuclear laboratory constraints just produces a
maximum mass about 1.7$M_\odot$ for the neutron star without
hyperons, whereas the hyperonization can further reduce the maximum
mass to a value below $1.4M_\odot$. In this case, the role of the
U-boson is constructive in increasing the maximum mass of neutron
stars, either as the EOS is softened by the creation of new degrees
of freedom, or the EOS is too soft to obtain a large maximum mass.

On the other hand, the radius of neutron stars is primarily
determined by the EOS  in the lower density region of $1\rho_0$ to
$2\rho_0$, see Refs.\cite{lat00,Li08} and references therein. Because
the symmetry energy in this density region offers the most important
ingredient of the pressure in pure neutron matter, the density
dependence of the symmetry energy plays a crucial role in determining
the radius of neutron stars. While in the present case the pressure
in the lower density region is increased appreciably  by the U-boson,
it is not surprising that the sensitive variation of the neutron star
radius is obtained accordingly. This is similar to the
non-relativistic case in Ref.~\cite{Wen09}. In fact, the radius of
neutron stars relies sensitively on the stiffness of the EOS. Thus,
the stiffening of the EOS caused by the U-boson gives rise to a
significant increase of the radius. Concretely, we can see from
Fig.~\ref{mr} that the larger rise of the radius comes up with the
more apparent stiffening role of the U-boson in softer models. It is
known that the radius of neutron stars extracted from the observation
can have a wide range due to the uncertainties of the distance
measurement and theoretical models used for the spectrum
analyses~\cite{lat00,Ha01,Lib06,Zh07}.  A more precise extraction of
the neutron star radius, probably through the coincident
measurements, thus becomes very significant,  because it can test the
non-Newtonian gravity due to its promising sensitivity to the star
radius.

\begin{figure}[thb]
\begin{center}
\vspace*{-5mm}
\includegraphics[height=7cm,width=8.5cm]{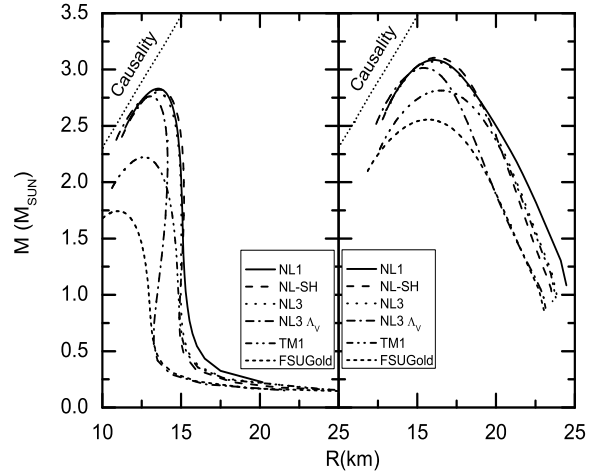}
 \end{center}
\vspace*{-5mm}\caption{Mass-radius relations  for various models with
the $(g_u / m_u)^2=0 GeV^{-2}$ (left panel) and the $(g_u /
m_u)^2=100 GeV^{-2}$ (right panel).\label{mr100}}
\end{figure}

To stress the role of the U-boson in the maximum mass and radius of
neutron stars, we depict in Fig.~\ref{mr100} the M-R relation for
various models with and without the U-boson. Here, for the case with
the inclusion of the U-boson, the calculation is performed with $(g_u
/ m_u)^2=100 GeV^{-2}$.  It is seen clearly that the large difference
in maximum masses with various types of  models can be reduced
largely by the U-boson with suitable parameter $(g_u / m_u)^2$. We
can see once again that the reduction of the difference is mainly
attributed to the role of the U-boson in the models featuring much
softer EOS's. Interestingly, we see that the uncertainty of the
radius for a canonical neutron star (with the mass $1.4M_\odot$) can
also be reduced by the U-boson.

In view of interesting and significant roles of the U-boson, we may
say that the task to look for the U-boson and further confirm the
non-Newtonian gravity is also confronted. The recent experimental
constraints on the relationship between parameters $\alpha$ ($g_u$)
and $\lambda$ ($m_u$) can be found in Ref.~\cite{Kr09}.  To recover
the stability of neutron stars using the EOS constrained by the
FOPI/GSI data~\cite{Xiao09}, the ratio $(g_u/m_u)^2\sim 100GeV^{-2}$
was found to be needed~\cite{Wen09}. In this work, the effect of the
U-boson is investigated within the parameter region
$(g_u/m_u)^2=0\sim 100GeV^{-2}$. To avoid the visible effect beyond
low energy constraints in finite nuclei, with these values of the
ratio parameter we may estimate that the mass of the U-boson should
be of order below $1 MeV$ with the coupling strength being almost or
at least three orders less than the fine-structure constant, while
these estimated orders can be compatible with parameter regions
allowed by a few experimental constraints, see Ref.~\cite{Kr09}. We
expect that more precision experiments will be performed to better
determine or exclude the parameter regions for the non-Newtonian
gravity.

At last, it is interesting to discuss the relevance between the
parameters of the non-Newtonian gravity touched upon in this work and
the solution to the dark matter problem. In order to explain the
flatness of the rotational curve of galactic spirals, one needs to
assume the non-luminous dark matter being the additional
gravitational source. Alternatively, the Newtonian gravity that was
well tested in the solar system may be assumed to fail at the large
distance scales of galaxies, and hence the Newtonian gravity should
be modified to be the non-Newtonian one~\cite{Man06}. The Yukawa-type
modification to the Newtonian gravity due to the boson exchange may
possibly be considered as a candidate to solve the dark matter
problem. In this work, the vector coupling of the U-boson that is
restrained by the U(1) symmetry produces a repulsion other than the
anticipated attraction. We may thus suppose to solve the dark matter
problem through the introduction of light scalar bosons. However,
since the flatness of the rotational curve requires a supplemental
force roughly linear inversely in the distance from the center of the
galaxy, even if the light scalar boson is assumed to provide the
needed attraction in one region, the exponential suppression factor
of the Yukawa-type potential (see Eq.(\ref{grav})) actually inhibits
the reproduction of the rotational curve in other regions. In deed,
in addition to the introduction of the light scalar boson, more
considerations are necessary to solve the dark matter
problem~\cite{Mb04}. On the other hand, we may explore the
constraints from the effect of the U-boson on the dark matter.
However, the coupling of the U-boson with the dark matter candidates
should be assumed to be much stronger than that with the normal
particles to explain the $511keV$ $\gamma$-ray observation while
simultaneously compatible with the low-energy
constraints~\cite{Bo04,Boe04,Bor06,Zhu07}. To sum up, we are
presently not able to restrain the parameters of the non-Newtonian
gravity originated from the U-boson exchange in this work directly by
using the effect of the U-boson on the dark matter and/or the
solution to the dark matter problem  with the modified Newtonian
dynamics. Nonetheless, this deserves further exploration. For
instance, the further first-principle understanding of the underlying
origin of the difference in the U-boson couplings to normal and dark
matter particles may open possibility to extract constraints on the
parameters of the non-Newtonian gravity.

\section{Summary}
\label{summary}

We have studied in this work the effects of the U-boson in RMF models
on the equation of state and subsequently the consequence in neutron
stars. All RMF models are chosen to have similarly nice reproduction
of saturation properties and ground-state properties of finite
nuclei, whereas they can give rise to a significantly large
difference in EOS's at high densities and mass-radius relations of
neutron stars. Interestingly, we find that the U-boson in models with
much softer EOS plays a much more significant role in increasing the
maximum mass of neutron stars. The distinction can be attributed
analytically to the different modification caused by the U-boson in
soft and stiff models to the pressure. Thus, the inclusion of the
U-boson may allow the existence of the non-nucleonic degrees of
freedom in the interior of large mass neutron stars initiated with
the favorably soft EOS of normal nuclear matter.  In addition, it is
worth notifying that the radius of canonical neutron stars in all
models can be sensitively modified by the U-boson due to its
stiffening role in the EOS.  Meanwhile, the difference in the
mass-radius relations predicted by various models can favorably be
reduced by increasing the coupling strength between the U-boson and
baryons. At last, constraints on the parameters of the non-Newtonian
gravity are discussed. Presently, we have not found the direct
relevance between the parameters of the non-Newtonian gravity
originated from the U-boson exchange and its effect on the dark
matter concerning the dark matter problem. Together with the future
coincident measurements and more precise extraction of the mass and
radius of neutron stars, the sensitive role of the U-boson in the M-R
relation may be helpfully used  to test the physics beyond the
standard model and consequently the existence of the non-Newtonian
gravity in the dense neutron star.

\section*{Acknowledgement}

Authors thank Professors De-Hua Wen, Lie-Wen Chen and Bao-An Li for
useful discussions. The work was supported in part by the SRTP Grant
of the Educational Ministry of China, the National Natural Science
Foundation of China under Grant No. 10975033, the China Jiangsu
Provincial Natural Science Foundation under Grant No.BK2009261, and
the China Major State Basic Research Development Program under
Contract No. 2007CB815004.

\end{document}